%% file: phonon_tunneling_ms_v33.tex
\documentclass[aps,showpacs,showkeys,twocolumn,aps]{revtex4}
\usepackage{makeidx}
\usepackage{amsmath}
\usepackage{eurosym}
\usepackage{amssymb}
\usepackage{graphicx}
\usepackage{dcolumn}
\usepackage{bm}
\usepackage{rotating}

\input{tcilatex}

\begin{document}

\title{Acoustic Phonon Tunneling and Heat Transport due to Evanescent
Electric Fields }
\author{Mika Prunnila}
\email{mika.prunnila@vtt.fi}
\author{Johanna Meltaus}
\affiliation{VTT Technical Research Centre of Finland, P.O.Box 1208, FIN-02044 VTT,
Espoo, Finland}
\affiliation{}
\date{\today }

\begin{abstract}
The authors describe how acoustic phonons can directly tunnel through vacuum
and, therefore, transmit energy and conduct heat between bodies that are
separated by a vacuum gap. This effect is enabled by introducing a coupling
mechanism, such as piezoelectricity, that strongly couples electric field
and lattice deformation. The electric field leaks into the vacuum as an
evanescent field, which leads to finite solid-vacuum-solid transmission
probability. Due to strong resonances in the system some phonons can go
through the vacuum gap with (or close to) unity transmission, which leads to
significant thermal conductance and heat flux.
\end{abstract}

\maketitle

The heat flux of thermally excited photons from an ideal black body at a
temperature $T$ is given by the Stefan-Boltzmann law, which states that the
flux is proportional to $T^{4}$. The net flux between two black bodies,
which are at different temperature and separated by a large vacuum gap $d$,
is the difference of their individual Stefan-Boltzmann fluxes. When a
realistic emissivity/absorption is considered this $T^{4}$ power law is
altered, but the essential physics still remains the same. However, when $d$
is smaller than the characteristic wave length $\lambda _{T}$ of thermal
spectrum various near-field radiation effects start to play crucial role in
the inter-body heat transport and new physics emerges (see Refs. \cite%
{joulain:2005,volokitin_rev:2007} for a review). In this $q_{T}d$ $<1$ limit
($q_{T}=2\pi /\lambda _{T}$ being the thermal wave vector) the heat flux is
enhanced particularly by evanescent waves, as explained correctly first by
Polder and Van Hove \cite{polder:1971}.\ Recent advances in experimental
techniques have enabled near-field heat transfer measurements from $\mu $m
down to 10 nm body distances.\cite%
{kittel:2005,narayanaswamy:2008,rousseau:2009} In this Letter, we propose
that at such distances a new type of evanescent field heat transfer
mechanism due to acoustic phonons can exist. 
\begin{figure*}[t]
\begin{center}
\includegraphics[width=175mm,height=!]{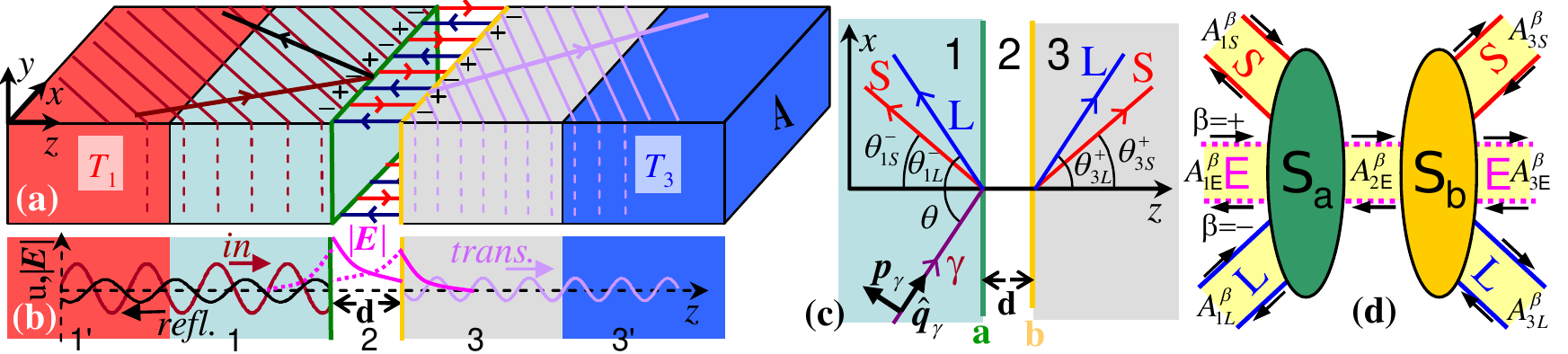}
\end{center}
\par
\caption{(Color online) Illustration of the phonon tunneling effect caused
by evanescent electric fields. (a) Cross-sectional slice of area $A$
illustrating a phonon incoming from a thermal path 1' at temperature $T_{1}$
and hitting a solid-vacuum interface. The phonon carries an electric field,
illustrated by +/- signs of polarization in between the wave fronts. The
polarization induces an electric field into the vacuum gap $2$ (some field
lines illustrated). The field enables finite transmission over the gap into
the thermal bath $3^{\prime }$ (at temperature $T_{3}$). The wave fronts of
the reflected phonon are not shown. (b) A projection along the z-axis
showing the spatial behaviour of the phonon waves ($u$) and of the
evanescent electric field ($\boldsymbol{E=E}_{E}$). The dashed curves depict
the "reflected" evanescent field. (c) Illustration of the scattering of the
incoming mode $\protect\gamma $ into different propagating modes ($L$%
=longitudinal and $S$=transversal). $\widehat{\boldsymbol{q}}_{\protect%
\gamma }$ and $\boldsymbol{p}_{\protect\gamma }$ are the propagation and
polarization vectors, respectively. (d) The scattering matrix formulation of
the problem (See Eq. \protect\ref{eq:S_and_amplitudes}). }
\label{fig:piezo_schem}
\end{figure*}

Even though acoustic phonons are the major heat carriers in dielectrics,
their effect on heat transfer through a vacuum gap has been considered to be
negligible, because they couple weakly to photons. Here we demonstrate by
theoretical means that significant energy transmission and heat flux is
possible if the acoustic phonons can induce an electric field, which then
can leak into the vacuum [see Figs. \ref{fig:piezo_schem}(a) and (b)]. Such
a mechanism is provided, for example, by the density response of free
carriers due to phonons or by the piezoelectric (PE) effect. Here we shall
focus on the latter, which gives rise to a strong coupling between phonon
induced material deformation and macroscopic electric fields. The
solid-vacuum-solid transmission phenomenon described here can be thought of
as an acoustic phonon tunneling through vacuum.

We consider a system in which two phonon black bodies $1\prime $ and $%
3^{\prime }$, which serve as fully thermalizing reservoirs, are connected to
bodies $1$ and $3$, which serve as wave guides for the propagating acoustic
phonons [Fig. \ref{fig:piezo_schem}(a)]. The propagating modes hit the
solid-vacuum interfaces and produce the evanescent electric fields of
interest [Fig. \ref{fig:piezo_schem}(b)], which lead to non-zero
solid-vacuum-solid energy transmission coefficient $\mathcal{T}_{\gamma
}=\sum\nolimits_{\mu }\mathcal{T}_{\mu \gamma }$ for incident mode $\gamma
=L,S$. Mode indices $L$ and$\ S$ stand for longitudinal and transversal,
respectively (the two transversal modes are not written explicitly). The
energy transmission probability $\mathcal{T}_{\mu \gamma }=\mathcal{T}_{\mu
\gamma }(qd,\theta )$ (from mode $\gamma $ to mode $\mu $) is a function of
the phonon polarizations, the absolute value of the incident phonon
wavevector $q$ and the angle of incidence $\theta $. Now, following Ref. 
\cite{swartz_rmp:1989} the thermal boundary conductance $G_{\gamma }$
arising from transmission of mode $\gamma $ can be defined as 
\begin{subequations}%
\label{eq:G_definition} 
\begin{eqnarray}
G_{\gamma } &=&\dint\limits_{v_{z}>0}\frac{d^{3}q}{\left( 2\pi \right) ^{3}}%
\hbar \omega _{\boldsymbol{q}}\frac{\partial N(\omega _{\boldsymbol{q}},T)}{%
\partial T}v_{\gamma }\mathcal{T}_{\gamma }  \notag \\
&=&\frac{2\pi ^{2}}{30}v_{\gamma }k_{B}q_{T}^{3}\left( 1+\frac{q_{T}}{4}%
\frac{\partial }{\partial q_{T}}\right) \mathcal{T}_{\gamma }^{eff}
\label{eq:G_T_eff} \\
\mathcal{T}_{\gamma }^{eff} &=&\frac{15}{\pi ^{4}}\left\langle
\int\limits_{0}^{q_{c}/q_{_{T}}}dx\frac{x^{3}}{e^{x}-1}\mathcal{T}_{\gamma
}\right\rangle ,  \label{eq:T_eff}
\end{eqnarray}%
\end{subequations}%
where $\mathcal{T}_{\gamma }^{eff}$ is the effective energy transmission
coefficient. The total thermal conductance and net heat flux are given by $%
G=\sum\nolimits_{\gamma }G_{\gamma }$ and $P=\tint_{T_{3}}^{T_{1}}GdT$,
respectively. Temperature dependency arises from phonon occupation number $%
N(\omega _{\boldsymbol{q}},T)=[\exp (\hbar \omega _{\boldsymbol{q}%
}/k_{B}T)-1]^{-1}$ and in $\mathcal{T}_{\gamma }^{eff}$\ the temperature is
buried into the thermal wave vector $q_{T}=2\pi /\lambda _{T}=\ k_{B}T/\hbar
v_{\gamma }$ ($\lambda _{T}$ is the thermal phonon wave length and $%
v_{\gamma }$ is the phonon velocity). For the sake of simplicity we have
assumed a linear dispersion $\omega _{\boldsymbol{q}}=v_{\gamma }q$ in Eq. (%
\ref{eq:T_eff}). The parameter $q_{c}\sim 1/a$ is the Brillouin/Debye
cut-off ($a$ being the lattice constant).\ The bracket $\left\langle \cdot
\cdot \cdot \right\rangle $ stands for a solid angle average over
half-space. Note that if $\mathcal{T}_{\gamma }=1$ and $\frac{q_{c}}{q_{T}}%
\rightarrow \infty $, then $\mathcal{T}_{\gamma }^{eff}=1$ and $P$ is equal
to the phonon black-body flux.

The energy transmissions are calculated from the scattering matrices
(S-matrix) $\mathcal{S}_{i}$ ($i=a,b$) of the two solid-vacuum interfaces $a$
and $b$. We define $\mathcal{S}_{i}$ in such a way that it couples the
amplitudes $A_{\alpha \nu }^{\beta }$ of propagating fields and
exponentially decaying near-fields [see Fig. \ref{fig:piezo_schem}(d)].
Here, $\beta =\pm $, where $+$ ($-$) refers to left-to-right
(right-to--left) propagating or decaying wave, $\alpha =1,2,3$ is the
material index (2 referring to the vacuum gap), and $\nu $ is the
mode/channel index ($\nu =E$ refers to the evanescent channel). For example,
for the interface $a$ we write 
\begin{equation}
\left( 
\begin{array}{c}
A_{1L}^{-} \\ 
A_{1S}^{-} \\ 
A_{1E}^{-} \\ 
A_{2E}^{+}%
\end{array}%
\right) =\mathcal{S}_{a}\left( 
\begin{array}{c}
A_{1\gamma }^{+} \\ 
A_{2E}^{-}%
\end{array}%
\right) =\left( 
\begin{array}{cc}
\underset{3\times 1}{r_{a}} & \underset{3\times 1}{t_{a}^{\prime }} \\ 
\underset{1\times 1}{t_{a}} & \underset{1\times 1}{r_{a}^{\prime }}%
\end{array}%
\right) \left( 
\begin{array}{c}
A_{1\gamma }^{+} \\ 
A_{2E}^{-}%
\end{array}%
\right) .  \label{eq:S_and_amplitudes}
\end{equation}%
The labels of the sub-matrices $r_{a}$, $t_{a}^{\prime }$, $t_{a}$ and $%
r_{a}^{\prime }$ indicate the size of these matrices. We assume that in all
materials there is only one channel that arises from evanescent electric
fields. This fully covers the case in which we neglect all retardation
effects and utilize the quasistatic approximation. Now, $A_{\alpha E}^{\beta
}$ is more conveniently related to the evanescent potential $\Phi _{E}$
instead of the evanescent electric field $\boldsymbol{E}_{E}$ ($\boldsymbol{E%
}_{E}=-\nabla \Phi _{E}$). The amplitudes $A_{\alpha E}^{\beta }$ are
coupled to the acoustic amplitudes if an acoustic phonon creates a periodic
charge density or polarization, which then creates an evanescent
field/potential due to a boundary. The retardation effects become important
if the oscillation period of the surface polarization (arising from the
acoustic phonons) is of the order of the time it takes light to make a round
trip across the gap. This leads to cut-off energy $\mathcal{E}%
_{c}=hc/2d\approx 620$ meV$\times (\mu $m$/d)$, below which our model is
valid. Note that $\mathcal{E}_{c}$ is above acoustic phonon energies when $%
d<10$ $\mu $m.

For the full solid-vacuum-solid system we need to find the total S-matrix $%
\mathcal{S}=\mathcal{S}_{a}\otimes \mathcal{S}_{b}$ that couples the
amplitudes of the different solids. From $\mathcal{S}$ we specifically need
the sub-matrix $t$, which describes the amplitude transmission. By solving $%
\mathcal{S}=\mathcal{S}_{a}\otimes \mathcal{S}_{b}$ and taking into account
the exponential factors arising from the finite distance $d$ we find%
\begin{equation}
t_{\mu \gamma }=\left\{ t_{b}^{\prime }\right\} _{\mu }\left[ 1-e^{-2\eta
qd}r_{a}^{\prime }r_{b}^{\prime }\right] ^{-1}\left\{ t_{a}\right\} _{\gamma
}e^{-\eta qd},  \label{eq:t_tot}
\end{equation}%
where we have inverted the z-axis for $\mathcal{S}_{b}$ so that $\mathcal{S}%
_{a}$ and $\mathcal{S}_{b}$ have a similar structure. Here the evanescent
field is excited by the oscillating polarization perpendicular to the z-axis
and, therefore, we have $\eta =\left\vert \sin \theta \right\vert $. The
energy transmission coefficients are given by $\mathcal{T}_{\mu \gamma
}=\alpha _{\mu \gamma }\left\vert t_{\mu \gamma }\right\vert ^{2}$, where $%
\alpha _{\mu \gamma }$ is a factor that converts the amplitude transmission
into energy transmission probability. We write the total energy transmission 
$\mathcal{T}_{\gamma }=\sum\nolimits_{\mu }\mathcal{T}_{\mu \gamma }$ in the
form 
\begin{equation}
\mathcal{T}_{\gamma }=\frac{e^{2\eta qd}}{(e^{2\eta qd}-R)^{2}+I^{2}}%
\sum\limits_{\mu }\mathcal{A}_{\mu \gamma },  \label{eq:t_energy}
\end{equation}%
where $A_{\mu \gamma }=\alpha _{\mu \gamma }\left\vert \left\{ t_{b}^{\prime
}\right\} _{\mu }\left\{ t_{a}\right\} _{\gamma }\right\vert ^{2}$, $R=\text{%
Re}\{r_{a}^{\prime }r_{b}^{\prime }\}$ and $I=\text{Im}\{r_{a}^{\prime
}r_{b}^{\prime }\}$. $\alpha _{\mu \gamma }$ can be determined from the
acoustic Poynting vector \cite{auld:b1} and in the simplest case of
isotropic solid we have $\alpha _{\mu \gamma }=\frac{\rho _{3}v_{\mu }}{\rho
_{1}v_{\gamma }}\text{Re}\left\{ \widehat{q}_{\mu }\right\} _{z},$where $%
\rho _{i}$ is the mass density of material $i$ and $\left\{ \widehat{q}_{\mu
}\right\} _{z}=\left( 1-(v_{\mu }^{2}/v_{\gamma }^{2})\sin ^{2}\theta
\right) ^{1/2}\ $\ is the $z$-direction propagation vector.

Our next task is to calculate the energy transmission coefficient $\mathcal{T%
}_{\gamma }$ in the case of two similar PE\ crystals separated by a vacuum
gap. In such a piezoacoustic system, we are dealing with coupled acoustic
and electromagnetic fields. Within the quasistatic model, the relevant field
quantities (see Table \ref{tab:fields}) 
\begin{table}[t]
\caption{Field quantities and constants.}
\label{tab:fields}%
\begin{tabular}{ccl}
\hline\hline
Symbol & Size & \ \ Name \\ \hline
$\boldsymbol{T}$
& $6\times 1$ & \ \ Stress \\ 
$\widehat{e}$ & $3\times 6$ & \ \ Piezo tensor \\ 
$\Phi $ & $1\times 1$ & \ \ Potential \\ 
$\widehat{c}$ & $6\times 6$ & \ \ Stiffness tensor \\ 
$\boldsymbol{u}$
& $3\times 1$ & \ \ Lattice displacement \\ 
$\boldsymbol{D}$
& $3\times 1$ & \ \ Electric displacement \\ 
$\widehat{\varepsilon }$ & $3\times 3$ & \ \ Dielectric constant%
\end{tabular}%
\end{table}
and their couplings are defined by \cite{auld:b1} 
\begin{subequations}%
\label{eq:fields} 
\begin{eqnarray}
\boldsymbol{T} &=&\widehat{e}^{T}\nabla \Phi +\widehat{c}\nabla _{u}%
\boldsymbol{u} \\
\boldsymbol{D} &=&-\widehat{\varepsilon }\nabla \Phi +\widehat{e}\nabla _{u}%
\boldsymbol{u} \\
\nabla ^{T}\widehat{\varepsilon }\nabla \Phi &=&\nabla ^{T}\widehat{e}\nabla
_{u}\boldsymbol{u},  \label{eq:piezo_poisson}
\end{eqnarray}%
\end{subequations}%
where $\nabla =[\partial /\partial x,\partial /\partial y,\partial /\partial
z]^{T}$ and $\nabla _{u}$ is the displacement-to-strain operator ($6\times 3$
matrix). The form of the operator $\nabla _{u}$ follows from the relation $%
\epsilon _{\alpha \beta }=\frac{1}{2}(\partial u_{\alpha }/\partial \beta
+\partial u_{\beta }/\partial \alpha )$ ($\alpha ,\beta =x,y,z$), which
couples the six symmetric strain components $\epsilon _{\alpha \beta }$ to
the components of displacement $\boldsymbol{u}$. Note that Eq. (\ref%
{eq:piezo_poisson}) is basically a Poisson equation and the source term $%
\nabla ^{T}\widehat{e}\nabla _{u}\boldsymbol{u}$ gives rise to the phonon
induced evanescent electric fields. The S-matrices $\mathcal{S}_{i}$ are
solved from a boundary condition equation that is obtained by requiring the
continuity of the potential $\Phi $, the normal component of electric
displacement $\boldsymbol{D}$ and the normal component of stress $%
\boldsymbol{T}$ at the interfaces. The boundary condition equation and the
resulting $\mathcal{S}_{i}$ are given in the supplementary material \cite%
{phon_tun_sup_mat:2010}. We calculate $\mathcal{T}_{\gamma }$ [Eq. (\ref%
{eq:t_energy})] by solving the S-matrix numerically. We adopt material
parameters that are close to that of ZnO \cite{auld:b1} with the simplifying
approximation $\left\{ \widehat{e}\right\} _{ij}=$ $\delta _{3j}\delta
_{i3}e_{33}$, where $e_{33}=1.3$ C/m$^{2}$. Furthermore, we assume that the
acoustic properties are isotropic and that the PE stiffening can be
neglected for the propagating modes. These assumptions have very little
effect on the angular averaged quantities of interest ($\mathcal{T}_{\gamma
}^{eff}$ and $G_{\gamma }$). We use isotropic values $c_{11}=209.7\times
10^{9}$ N/m$^{2},c_{44}=42\times 10^{9}$ N/m$^{2}$ and $\left\{ \widehat{%
\varepsilon }\right\} _{ij}=10\varepsilon _{0}\delta _{ij}$. The phonon
velocities are given by $v_{L}=\sqrt{c_{11}/\rho }=\allowbreak
6119.\,\allowbreak 3$ m/s and $v_{S}=\sqrt{c_{44}/\rho }=2738.\,6$ m/s,
where the mass density $\rho =5600$ kg/m$^{3}$.

\begin{figure}[t]
\begin{center}
\includegraphics[width=83mm,height=!]{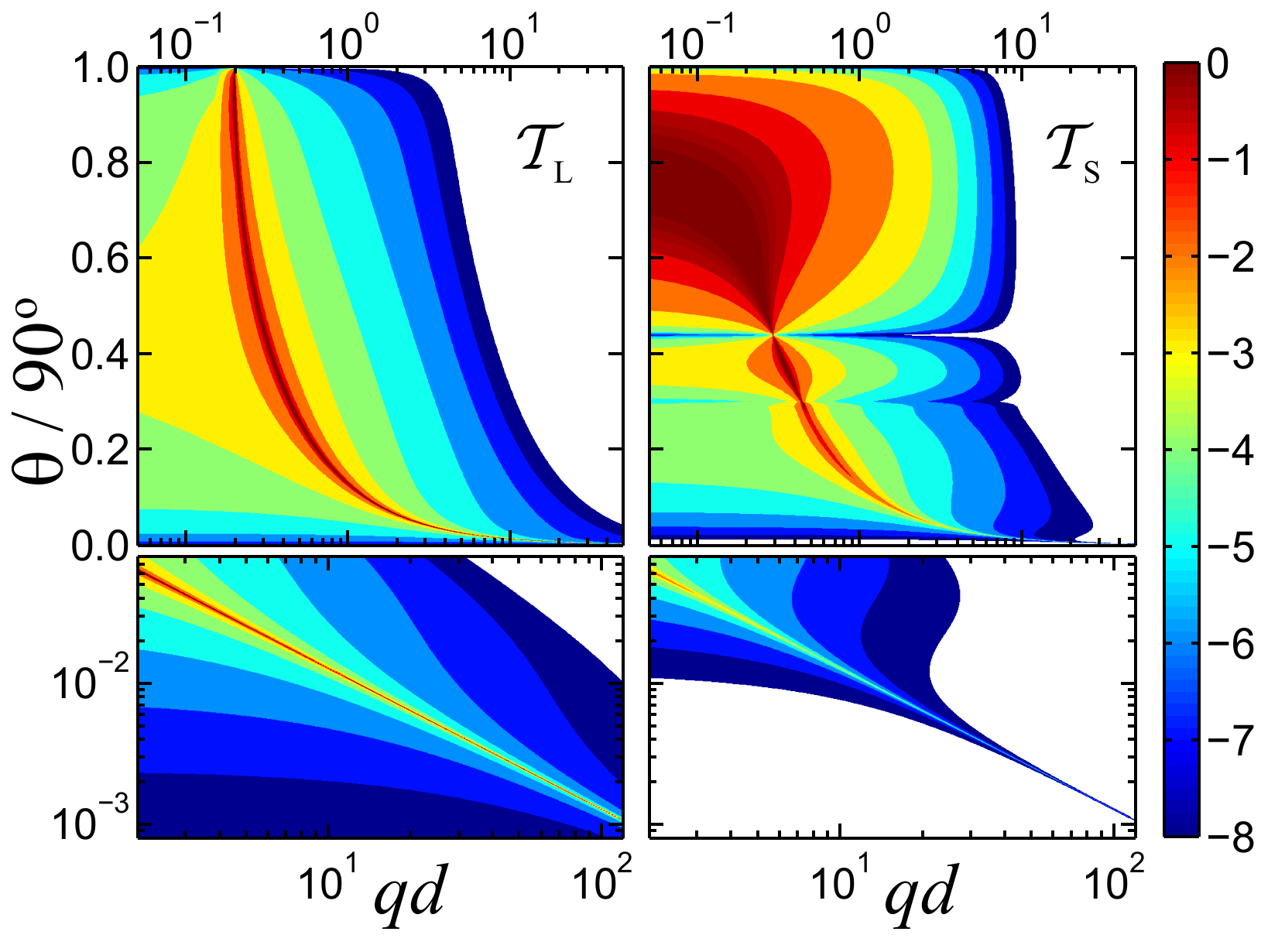}
\end{center}
\par
\caption{(Color online) Logarithmic contour plot of energy transmission
coefficients $\mathcal{T}_{\protect\gamma }$ ($\protect\gamma =L,S$) between
two PE bodies as a function of normalized wave vector $qd$ and the angle of
incidence $\protect\theta $. White regions have $\mathcal{T}_{\protect\gamma %
}<10^{-8}$. The lower panels show log-log blow-up of the small-$\protect%
\theta $ large-$qd$ region. }
\label{fig:transmission}
\end{figure}

Figure \ref{fig:transmission} shows $\mathcal{T}_{\gamma }(qd,\theta )$ for
phonon tunneling from a PE material to another across a vacuum gap. 
There are two local minima in $\mathcal{T}_{S}$: one at $\theta \approx
25^{\circ }$ and another one at $\theta \approx 40^{\circ }$. The position
of the latter minima can be related to the phase change of the evanescent
electric field in the vacuum-solid reflection. The former minima is the
critical incidence angle where the quasi-longitudinal surface mode is
exited. Most important effects for the energy transmission are the strong
resonance features for both modes: for some $(qd,\theta )$-values $\mathcal{T%
}_{\gamma }$ is very large or even equal to unity. The resonances arise from
the multiple reflections of the evanescent field in the vacuum gap, which
leads to the $\left[ (e^{2\eta qd}-R)^{2}+I^{2}\right] ^{-1}$ factor in $%
\mathcal{T}_{\gamma }$ [see Eq. (\ref{eq:t_energy})]. The elements $%
r_{a,b}^{\prime }$ are dictated by the vacuum-solid boundary conditions and
their real parts or moduli are not limited below unity. Thus we may have $R>1
$ and $I\ll 1$, which leads to a sharp resonance peak at $2\eta qd=\ln R$.
Indeed, for $\gamma =L,$ we have $R\sim 1.5$ for all incident angles and
this is the origin of the sharp maximum trajectory in $\mathcal{T}%
_{L}(qd,\theta )$. For $\gamma =S$, $R>1$ only for angles $\theta <68^{\circ
}$, and above this threshold no sharp resonances exist. For small $\theta $
at $qd\gg 1$, $\mathcal{T}_{L}=1$ at the resonance, whereas the amplitude of
the resonance peak in $\mathcal{T}_{S}$\ decays as a function of $qd$ (see
the lower panels of Fig. \ref{fig:transmission}). Note that the phonon
resonant tunneling here has a striking similarity to the resonant photon
tunneling described, for example, in Ref. \cite{volokitin_rev:2007}.

Inserting the calculated $\mathcal{T}_{\gamma }$ into Eq. (\ref%
{eq:G_definition}) and performing numerical integration, we obtain $\mathcal{%
T}_{\gamma }^{eff}$ and $G_{\gamma }$. The results are presented in Fig. \ref%
{fig:G}. 
\begin{figure}[t]
\begin{center}
\includegraphics[width=67mm,height=!]{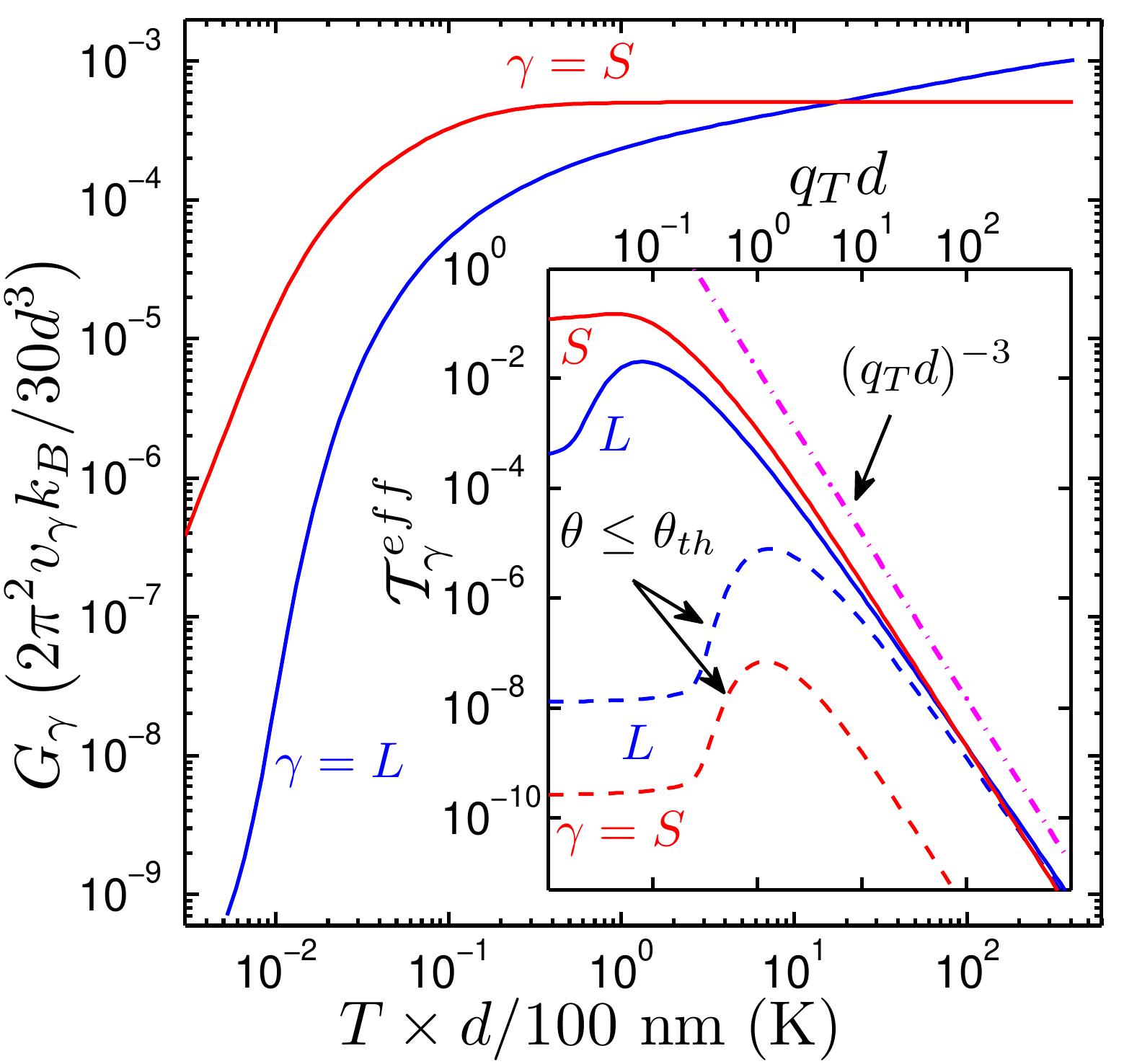}
\end{center}
\caption{(Color online) Heat transport between two PE bodies: effective
transmission $\mathcal{T}_{\protect\gamma }^{eff}$ (the inset) as a function
of $q_{T}d$ and the interface thermal conductance $G_{\protect\gamma }$ (the
main part) as a function of temperature. The curves are obtained from the $%
\mathcal{T}_{\protect\gamma }$ of Fig. (\protect\ref{fig:transmission}) and
Eq. (\protect\ref{eq:G_definition}) (we have set $q_{c}/q_{T}\rightarrow
\infty $ ). The dashed curves in the inset are small angle contributions $%
\mathcal{T}_{\protect\gamma ,\protect\theta <\protect\theta _{th}}^{eff}$
with $\protect\theta _{th}=3^{\circ }$. The dot-dash curve identicates $%
\left( q_{T}d\right) ^{-3}$ slope. For the y-axis units of the main Figure
we have $2\protect\pi ^{2}v_{\protect\gamma }k_{B}/30d^{3}=$ $Y_{\protect%
\gamma }\times (100$ nm$/d)^{3}$ W/Km$^{2}$, where $Y_{L(S)}=55.6$ $\ $($%
24.9 $).}
\label{fig:G}
\end{figure}
We observe that at low temperatures when $q_{T}d\ll 1$ the effective
transmission $\mathcal{T}_{S}^{eff}$ is $\sim 15\%$ of the unity
transmission resulting in relatively large $G_{S}$, which is of the order of
the maximum possible thermal conductance allowed by Eq. (\ref{eq:G_T_eff}).
The large $\mathcal{T}_{S}^{eff}$ at low temperatures follows from the
broadened resonances and from the large transmission amplitude for the $S$%
-mode for large $\theta $ and small $qd$ (see Fig. \ref{fig:transmission}).
For $\gamma =L$ we have sharp resonances leading to a small effective
transmission and to $G_{L}\ll G_{S}$ at low temperatures.

Next, we want to investigate the dependency of $\mathcal{T}_{\gamma }^{eff}$
on $q_{T}d$ and the contribution of different phonons in the $(qd,\theta )$
phase space to the heat transport. We divide $\mathcal{T}_{\gamma }^{eff}$
into small angle $\mathcal{T}_{\gamma ,\theta <\theta _{th}}^{eff}$and large
angle $\mathcal{T}_{\gamma ,\theta >\theta _{th}}^{eff}$ contributions:%
\begin{equation}
\mathcal{T}_{\gamma }^{eff}=\mathcal{T}_{\gamma ,\theta <\theta _{th}}^{eff}+%
\mathcal{T}_{\gamma ,\theta >\theta _{th}}^{eff},  \label{eq:T_eff_SA_LA}
\end{equation}%
where the angular integrals of the different contributions are limited by
the threshold $\theta _{th}$, which is chosen in such away that major
contribution to $\mathcal{T}_{\gamma ,\theta <\theta _{th}}^{eff}$ comes
from the resonance trajectory at $qd>1$. The exact choice of $\theta _{th}$
is, therefore, somewhat arbitrary and in the numerical calculations we
define $\theta _{th}=3^{\circ }$, whence $\theta <\theta _{th}$ roughly
correspond to the phase space of the lower panels of Fig. \ref%
{fig:transmission}. At the high temperature limit ($q_{T}d\gg 1$), small and
large angle effective transmissions can be written as\cite%
{phon_tun_sup_mat:2010} 
\begin{subequations}%
\label{eq:T_eff_limits} 
\begin{eqnarray}
\mathcal{T}_{\gamma ,\theta <\theta _{th}}^{eff} &\approx &\dsum\limits_{\mu
}\frac{15}{\pi ^{3}}\frac{\left[ \ln R(0)\right] ^{n+1}}{\left(
2q_{T}d\right) ^{2+n}}  \notag \\
&&\times \frac{f_{\mu \gamma }^{n}(0)}{n!}F_{n}(\frac{q_{c}}{q_{_{T}}}%
,\theta _{th})  \label{eq:T_eff_SA} \\
\mathcal{T}_{\gamma ,\theta >\theta _{th}}^{eff} &\approx &\frac{15}{\pi ^{4}%
}\frac{1}{\left( q_{T}d\right) ^{3}}\left\langle
\int\limits_{0}^{q_{c}d}dyy^{2}\mathcal{T}_{\gamma }(y,\theta )\right\rangle
_{\theta \geq \theta _{th}},  \label{eq:T_eff_LA}
\end{eqnarray}%
\end{subequations}
where $F_{n}(\frac{q_{c}}{q_{_{T}}},\theta
_{th})=\tint\nolimits_{x_{th}}^{q_{c}/q_{_{T}}}dx\frac{x^{1-n}}{\exp (x)-1}$
with $x_{th}=\ln R(\theta _{th})/(2q_{T}d\sin \theta _{th})$, $f_{\mu \gamma
}(\theta )=\mathcal{A}_{\mu \gamma }(\theta )/I(\theta )$ and $f_{\mu \gamma
}^{n}(\theta )$ is the first non-zero derivative of $f_{\mu \gamma }(\theta
) $. Due to the decay of the resonance for the $S$-mode, $\mathcal{T}%
_{S,\theta <\theta _{th}}^{eff}$ is very small. This means that practically
all contribution to $\mathcal{T}_{S}^{eff}$ and $G_{S}$ at any temperature
comes from the large angle part $\mathcal{T}_{S,\theta >\theta _{th}}^{eff}$%
. At high temperatures, $\mathcal{T}_{\gamma ,\theta >\theta _{th}}^{eff}$
is given by Eq. (\ref{eq:T_eff_LA}), which explains the $\left(
q_{T}d\right) ^{-3}$ fall-off of $\mathcal{T}_{S}^{eff}$ and the saturation
of $G_{S}\rightarrow G_{S}^{sat}$ in Fig. \ref{fig:G}. The saturation value $%
G_{S}^{sat}=\frac{k_{B}v_{S}}{4\pi ^{2}d^{3}}\left\langle
\tint\nolimits_{0}^{q_{c}d}dyy^{2}\mathcal{T}_{\gamma }(y,\theta
)\right\rangle _{\theta \geq \theta _{th}}$ follows from Eqs. (\ref%
{eq:T_eff_LA}) and (\ref{eq:G_T_eff}). For the $L$-mode the high temperature 
$\mathcal{T}_{\gamma }^{eff}$ has also a large contribution arising from $%
\mathcal{T}_{L,\theta <\theta _{th}}^{eff}$. This contribution eventually
exceeds the large angle part and, as a result, there is no saturation in $%
G_{L}$. Close to $\theta =0,$ $\mathcal{T}_{L,\theta <\theta _{th}}^{eff}$is
dominated by $f_{LL}^{1}(\theta )$, and$\ $by setting $\frac{q_{c}}{q_{T}}%
\rightarrow \infty $ in Eq. (\ref{eq:T_eff_SA}), we find that $\mathcal{T}%
_{L,\theta <\theta _{th}}^{eff}\propto \left( q_{T}d\right)
^{-3}\sum\nolimits_{k=1}^{\infty }\frac{\exp (-kx_{th})}{k}$, resembling a $%
\left( q_{T}d\right) ^{-5/2}$ behavior. It should be noted that the
temperature dependency of the large angle contribution is not affected by
the ratio $q_{c}/q_{T}$. As $q_{c}^{-1}$ is of the order of the lattice
constant $a$, Eq. (\ref{eq:T_eff_LA}) is valid if $d>a$. The small angle
contribution depends on $q_{c}/q_{T}$. For example, if $q_{c}/q_{T}\ll 1$,
then $\mathcal{T}_{L,\theta <\theta _{th}}^{eff}\propto \left( q_{T}d\right)
^{-3}\ln \left[ \frac{2q_{c}d\sin \theta _{th}}{\ln R(\theta _{th})}\right] $%
. Thus, $G_{L}$ also saturates when the cut-off $q_{c}$ is exceeded.

At this point we summarize our findings:\ We have formulated the thermal
boundary conductance of a solid-vacuum-solid system [Eq. (\ref%
{eq:G_definition})] using an acoustic phonon energy transmission probability
[Eq. (\ref{eq:t_energy})] given by scattering matrix [Eq. (\ref%
{eq:S_and_amplitudes})], which couples the different solids by an evanescent
channel produced by phonon induced electric fields (Fig. \ref%
{fig:piezo_schem}). These fields can lead to a significant energy
transmission probability for some parts of the wave vector-angle of
incidence $(qd,\theta )$ phase space (Fig. \ref{fig:transmission}). The
transmission exhibits resonances, which arise from the multiple reflections
of the evanescent field in the vacuum gap. The $L$-mode shows a strong and
sharp resonance trajectory across the $(qd,\theta )$-plane with a unity
energy transmission probability. The resonance trajectory approaches $\theta
=0$ as $qd\rightarrow \infty $. The S-mode exhibits similar features.
However, for a small $qd$ and a large $\theta $ the resonance is broadened
and in the $qd\gg 1$ limit the energy transmission decays rapidly. At low
temperatures ($q_{T}d<1$), the broadened resonance at small $qd$ leads to a
large contribution from the $S$-mode to the thermal conductance ($\sim 15\%$
of the maximum possible thermal conductance defined by unity transmission).
The small $qd$ /large $\theta $ part of the phase space dominates the $S$%
-mode heat transport leading to a saturation of the thermal conductance of
the $S$-mode at high temperatures (Fig. \ref{fig:G}). If the thermal phonon
wave length $\lambda _{T}$ is sufficiently smaller than the lattice
constant, the $L$-mode shows no such saturation. This is due to the strong
resonance in the transmission that persists even at $qd\gg 1$ and,
therefore, allows short wave length phonons to go through the vacuum gap.
Note that our method to calculate the thermal conductance is essentially
similar to that adopted in (ballistic) thermal boundary resistance
calculations \cite{swartz_rmp:1989}, which means that the thermal excitation
of local acusto-electric modes related to the solid-vacuum boundaries is not
taken into account. Here, the boundary effects and the evanescent fields are
solely due to radiation of bulk acoustic phonons. One can also explicitly
take into account thermal excitation and couplings of the local modes,
expected to enhance the interbody coupling and the thermal conductance. This
approach, typically adopted in photon near-field studies\cite%
{joulain:2005,volokitin_rev:2007}, will be left for future investigations.

Acoustic phonon heat transfer through vacuum can be experimentally verified
with direct thermal conductance measurements, using piezoelectric materials.
The transmission of acoustic energy across a vacuum gap at long wave lengths
can be investigated by measuring the transmission of a single acoustic wave
by utilizing, for example, surface acoustic wave devices. Finally, we note
that the effects described in this paper can also contribute to the thermal
conductivity of polycrystalline piezoelectric materials and percolation
systems consisting of piezoelectric particles. Furthermore, as free
electrons couple strongly to electric fields, a similar heat transport
effect, that is discussed in this Letter, can also occur between metallic
and piezoelectric bodies.

\begin{acknowledgments}
We acknowledge useful discussions with P.-O. Chapuis, B. Djafari-Rouhani, J.
Ahopelto and T. Pensala. This work was financially supported by EC through
the project \#216176 NANOPACK.
\end{acknowledgments}

\bibliographystyle{apsrev}
\bibliography{books,notes,notes_phonon_tunneling,phonon_refs,photon_nf_refs}

\input{dummy.tex}%

\newpage

\input{phonon_tunneling_sup_material_v23.tex}%

\end{document}

%% file: dummy.tex

%% file: phonon_tunneling_sup_material_v23.tex
\begin{widetext}
\newpage
\end{widetext}
\begin{widetext}
\begin{center}
{\LARGE SUPPLEMENTARY MATERIAL}
\end{center}
\end{widetext}

\markright{Supplementary Material, EPAPS Document No. XXX}
\section{Derivation of the scattering matrix in a piezoacoustic solid-vacuum
problem}

In the quasi-static approximation, the piezoelectric constitutive relations
for stress $\boldsymbol{T}$, electric displacement $\boldsymbol{D}$ and
electric potential $\Phi $, generated by a displacement $\boldsymbol{u}$,
can be written as \cite{auld:b1}%
\begin{subequations}%
\label{eq:fields}

\begin{eqnarray}
\boldsymbol{T} &=&\widehat{e}^{T}\nabla \Phi +\widehat{c}\nabla _{u}%
\boldsymbol{u}  \label{eq:T-const} \\
\boldsymbol{D} &=&-\widehat{\varepsilon }\nabla \Phi +\widehat{e}\nabla _{u}%
\boldsymbol{u}  \label{eq:D-const} \\
\nabla ^{T}\hat{\varepsilon}\nabla \Phi &=&\nabla ^{T}\hat{e}\nabla _{u}%
\boldsymbol{u}.  \label{eq:Phi-const}
\end{eqnarray}%
\end{subequations}%
The electric field $\boldsymbol{E}$ is given by $\boldsymbol{E}=-\nabla \Phi 
$. Tensors $\hat{\varepsilon}$, $\hat{e}$ and $\hat{c}$ are dielectric
permittivity, piezoelectric coupling and stiffness tensors, respectively.
Above Equations are written following the abbreviated tensor subscript
notation \cite{auld:b1}. The field quantities are are summarized in the
Table I of the main article. Here, we have defined operators $\nabla $ and $%
\nabla _{u}$ according to 
\begin{subequations}%
\label{eq:nablas} 
\begin{gather}
\nabla ^{T}=%
\begin{bmatrix}
\partial /\partial x & \partial /\partial y & \partial /\partial z%
\end{bmatrix}
\\
\nabla _{u}^{T}=%
\begin{bmatrix}
\partial /\partial x & 0 & 0 & 0 & \partial /\partial z & \partial /\partial
y \\ 
0 & \partial /\partial y & 0 & \partial /\partial z & 0 & \partial /\partial
x \\ 
0 & 0 & \partial /\partial z & \partial /\partial y & \partial /\partial x & 
0%
\end{bmatrix}%
.
\end{gather}%
\end{subequations}%
For displacement $\boldsymbol{u}$ with a known wave vector $\boldsymbol{q}$,
amplitude $A$ and polarization vector $\boldsymbol{p}$ 
\begin{equation}
\boldsymbol{u}=A\boldsymbol{p}e^{-j\boldsymbol{q}\cdot \boldsymbol{r}},
\label{eq:plane_wave}
\end{equation}%
the above operators can be written as%
\begin{subequations}
\begin{eqnarray}
\nabla &=&-jq\boldsymbol{L} \\
\nabla _{u} &=&-jq\boldsymbol{L}_{u},
\end{eqnarray}%
\end{subequations}%
where $\boldsymbol{L}$ and $\boldsymbol{L}_{u}$ are defined with the
components $l_{x}$, $l_{y}$ and $l_{z}$ of the unit propagation vector
parallel to wave vector $\boldsymbol{q}$:%
\begin{subequations}
\begin{eqnarray}
\boldsymbol{L}^{T} &=&%
\begin{bmatrix}
l_{x} & l_{y} & l_{z}%
\end{bmatrix}
\\
\boldsymbol{L}_{u}^{T} &=&\left[ 
\begin{array}{cccccc}
l_{x} & 0 & 0 & 0 & l_{z} & l_{y} \\ 
0 & l_{y} & 0 & l_{z} & 0 & l_{x} \\ 
0 & 0 & l_{z} & l_{y} & l_{x} & 0%
\end{array}%
\right] .
\end{eqnarray}%
\end{subequations}%
Re-writing the constitutive relations for $\boldsymbol{u}=A\boldsymbol{p}%
e^{-j\boldsymbol{q}\cdot \boldsymbol{r}}$ with $\boldsymbol{L}$ and $%
\boldsymbol{L}_{u}$, we get 
\begin{subequations}%
\label{eq:const-rel} 
\begin{eqnarray}
\boldsymbol{T} &=&-jq\left[ \frac{\hat{e}^{T}\hat{e}}{\boldsymbol{L}^{T}\hat{%
\varepsilon}\boldsymbol{L}}+\hat{c}^{E}\right] \boldsymbol{L}_{u}\boldsymbol{%
p}  \label{eq:T-constL} \\
\boldsymbol{D} &=&jq\left[ \hat{\varepsilon}\frac{1}{\boldsymbol{L}^{T}\hat{%
\varepsilon}\boldsymbol{L}}\boldsymbol{LL}^{T}-1\right] \hat{e}\boldsymbol{L}%
_{u}\boldsymbol{p}  \label{eq:D-constL} \\
\Phi &=&\frac{\boldsymbol{L}^{T}\hat{e}\boldsymbol{L}_{u}\boldsymbol{p}}{%
\boldsymbol{L}^{T}\hat{\varepsilon}\boldsymbol{L}},
\end{eqnarray}%
\end{subequations}%
where we have excluded the amplitude-phase factor $Ae^{-j\boldsymbol{q}\cdot 
\boldsymbol{r}}$. The evanescent electric field is described through
evanescent potential $\Phi _{E}=A_{E}e^{-j\boldsymbol{q}_{\rho }\cdot 
\boldsymbol{r}+q_{e}z}$ ($\boldsymbol{q}_{\rho }=\boldsymbol{q}_{x}+%
\boldsymbol{q}_{y}$), for which similar field relations are given by 
\begin{subequations}%
\label{eq:pot-rel} 
\begin{eqnarray}
\boldsymbol{T}_{E} &=&e^{T}\nabla \Phi _{E}=-jqe^{T}\boldsymbol{L}_{E}
\label{eq:T-pot} \\
\boldsymbol{D}_{E} &=&-\hat{\varepsilon}\nabla \Phi _{E}=jq\hat{\varepsilon}%
\boldsymbol{L}_{E},  \label{eq:D-pot}
\end{eqnarray}%
\end{subequations}%
with%
\begin{equation}
\boldsymbol{L}_{E}^{T}=%
\begin{bmatrix}
l_{x} & l_{y} & l_{E}%
\end{bmatrix}%
,
\end{equation}%
where $l_{e}=q_{e}/jq=\pm \left\vert q_{\rho }\right\vert /jq=\pm 1/\sqrt{2}%
j $.

Boundary conditions at free piezoelectric boundary between materials 1 and 2
(vacuum) require the absence of stresses and the continuation of $\Phi $ and
the normal component of $\boldsymbol{D}$. Thus, we have boundary conditions%
\begin{subequations}%
\label{eq:BC} 
\begin{eqnarray}
\boldsymbol{T}_{1}\hat{n}_{1} &=&0  \label{eq:T-boundary} \\
\boldsymbol{D}_{1}\hat{n}_{1}^{T} &=&\boldsymbol{D}_{2}\hat{n}_{2}^{T}
\label{eq:D-boundary} \\
\Phi _{1} &=&\Phi _{2},  \label{eq:Phi-boundary}
\end{eqnarray}%
\end{subequations}%
where $\hat{n}_{\alpha }$ ($\alpha =1,2$ and $\left\vert \hat{n}_{\alpha
}\right\vert =1$) are the normal vectors of the boundary pointing into
material $\alpha $. Eq.~(\ref{eq:T-boundary}) involves tensor form of $%
\boldsymbol{T}$: 
\begin{equation}
\boldsymbol{T}=%
\begin{bmatrix}
T_{1} & T_{6} & T_{5} \\ 
T_{6} & T_{2} & T_{4} \\ 
T_{5} & T_{4} & T_{3}%
\end{bmatrix}%
.  \label{eq:T-tensor}
\end{equation}%
Now, we include the notation of the main article, which writes the
amplitudes as $A_{\alpha \nu }^{\beta }$, where $\beta =+(-)$ denotes the
left-to-right (right-to-left) propagating (or evanescent) wave, $\alpha $
denotes the material, and $\nu $ denotes the wave / field mode ($L$%
=longitudinal, $S$=shear, $E$=evanescent). From Eqs. (\ref{eq:const-rel}), (%
\ref{eq:pot-rel}) and (\ref{eq:BC}) we find the boundary condition matrix
equation 
\begin{eqnarray}
&&%
\begin{bmatrix}
\left[ \boldsymbol{T}_{1\gamma }^{+}\widehat{n}_{1}\right] & \left[ 
\boldsymbol{T}_{1L}^{-}\widehat{n}_{1}\right] & \left[ \boldsymbol{T}%
_{1S}^{-}\widehat{n}_{1}\right] & \left[ \boldsymbol{T}_{1E}^{-}\widehat{n}%
_{1}\right] \\ 
\boldsymbol{D}_{1\gamma }^{+}\widehat{n}_{1}^{T} & \boldsymbol{D}_{1L}^{-}%
\widehat{n}_{1}^{T} & \boldsymbol{D}_{1S}^{-}\widehat{n}_{1}^{T} & 
\boldsymbol{D}_{1E}^{-}\widehat{n}_{1}^{T} \\ 
\Phi _{1\gamma }^{+} & \Phi _{1L}^{-} & \Phi _{1S}^{-} & \Phi _{1E}^{-}%
\end{bmatrix}%
\begin{bmatrix}
A_{1\gamma }^{+} \\ 
A_{1L}^{-} \\ 
A_{1S}^{-} \\ 
A_{1E}^{-}%
\end{bmatrix}
\notag \\
&=&%
\begin{bmatrix}
\left[ 0\right] & \left[ 0\right] \\ 
\boldsymbol{D}_{2E}^{+}\widehat{n}_{2}^{T} & \boldsymbol{D}_{2E}^{-}\widehat{%
n}_{2}^{T} \\ 
\Phi _{2E}^{+} & \Phi _{2E}^{-}%
\end{bmatrix}%
\begin{bmatrix}
A_{2E}^{+} \\ 
A_{2E}^{-}%
\end{bmatrix}%
.  \label{eq:BC_matrix}
\end{eqnarray}%
Here, $A_{1\gamma }^{+}$ is the amplitude of an incident propagating mode $%
\gamma =L,S$ in material 1. The acoustic field related elements $\boldsymbol{%
T}_{\alpha \nu }^{\beta }$, $\boldsymbol{D}_{\alpha \nu }^{\beta }$, and $%
\Phi _{\alpha \nu }^{\beta }$($\nu =S,L$) are given by Eqs. (\ref%
{eq:const-rel}) and the evanescent field related elements $\boldsymbol{T}%
_{\alpha E}^{\beta }$ and $\boldsymbol{D}_{\alpha E}^{\beta }$ are give by
Eqs. (\ref{eq:pot-rel}). The elements $\Phi _{\alpha E}^{\beta }$ are
defined by the continuity of the potential and, therefore, $\Phi _{\alpha
E}^{\beta }=1$. The elements in square brackets (for example $\left[ 
\boldsymbol{T}_{1E}^{-}\widehat{n}_{1}\right] $) are not scalars: here $\dim
\left\{ \left[ x\right] \right\} =2\times 1$. From Eq. (\ref{eq:BC_matrix})
we obtain the scattering matrix equation 
\begin{equation}
\begin{bmatrix}
A_{1L}^{-} \\ 
A_{1S}^{-} \\ 
A_{1E}^{-} \\ 
A_{2E}^{+}%
\end{bmatrix}%
=\mathcal{S}%
\begin{bmatrix}
A_{1\gamma }^{+} \\ 
A_{2E}^{-}%
\end{bmatrix}%
,
\end{equation}%
where the scattering matrix $\mathcal{S}$ is defined by 
\begin{eqnarray}
\mathcal{S} &=&%
\begin{bmatrix}
\left[ \boldsymbol{T}_{1L}^{-}\widehat{n}_{1}\right] & \left[ \boldsymbol{T}%
_{1S}^{-}\widehat{n}_{1}\right] & \left[ \boldsymbol{T}_{1E}^{-}\widehat{n}%
_{1}\right] & \left[ 0\right] \\ 
\boldsymbol{D}_{1L}^{-}\widehat{n}_{1}^{T} & \boldsymbol{D}_{1S}^{-}\widehat{%
n}_{1}^{T} & \boldsymbol{D}_{1E}^{-}\widehat{n}_{1}^{T} & -\boldsymbol{D}%
_{2E}^{+}\widehat{n}_{2}^{T} \\ 
\Phi _{1L}^{-} & \Phi _{1S}^{-} & \Phi _{1E}^{-} & -\Phi _{2E}^{+}%
\end{bmatrix}%
^{-1}  \notag \\
&&\times 
\begin{bmatrix}
-\left[ \boldsymbol{T}_{1\gamma }^{+}\widehat{n}_{1}\right] & \left[ 0\right]
\\ 
-\boldsymbol{D}_{1\gamma }^{+}\widehat{n}_{1}^{T} & \boldsymbol{D}_{2E}^{-}%
\widehat{n}_{2}^{T} \\ 
-\Phi _{1\gamma }^{+} & \Phi _{2E}^{-}%
\end{bmatrix}%
.
\end{eqnarray}%
This equation is utilized in solving the S-matrix \ and the energy
transmission in the main article. Note that all $\boldsymbol{q}$-vectors $%
\boldsymbol{q}_{\alpha \nu }^{\beta }$ are proportional to the incident mode
wave number $q=\left\vert \boldsymbol{q}_{1\gamma }^{+}\right\vert $ and,
therefore, $\mathcal{S}$ does not depend on $q$. Furthermore, as incident
mode $\gamma $ is a parameter of $\mathcal{S}$ we must actually consider $%
\mathcal{S}$ as a sub-matrix of the total S-matrix.

\section{$\mathcal{T}_{\protect\gamma }^{eff}$ at $q_{T}d\gg 1$ limit}

In the main article we divide the effective transmission $\mathcal{T}%
_{\gamma }^{eff}$ into small angle $\mathcal{T}_{\gamma ,\theta <\theta
_{th}}^{eff}$and large angle $\mathcal{T}_{\gamma ,\theta >\theta
_{th}}^{eff}$ contributions:%
\begin{equation}
\mathcal{T}_{\gamma }^{eff}=\mathcal{T}_{\gamma ,\theta <\theta _{th}}^{eff}+%
\mathcal{T}_{\gamma ,\theta >\theta _{th}}^{eff},  \label{eq:T_eff_SA_LA}
\end{equation}%
where the angular integrals of the different contributions are limited by
the threshold $\theta _{th}$. We will describe the derivation of the high
temperature ($q_{T}d\gg 1$) analytical formulas of $\mathcal{T}_{\gamma
,\theta <\theta _{th}}^{eff}$ and $\mathcal{T}_{\gamma ,\theta >\theta
_{th}}^{eff}$ below.

\subsection{Small angle contribution $\mathcal{T}_{\protect\gamma ,\protect%
\theta <\protect\theta _{th}}^{eff}$}

The energy transmission coefficient is given by (see the main article)%
\begin{equation}
\mathcal{T}_{\gamma }=\frac{e^{2\eta qd}}{(e^{2\eta qd}-R)^{2}+I^{2}}%
\sum\limits_{\mu }\mathcal{A}_{\mu \gamma },
\end{equation}%
where $A_{\mu \gamma }=\alpha _{\mu \gamma }\left\vert \left\{ t_{b}^{\prime
}\right\} _{\mu }\left\{ t_{a}\right\} _{\gamma }\right\vert ^{2}$, $R=\text{%
Re}\{r_{a}^{\prime }r_{b}^{\prime }\}$ and $I=\text{Im}\{r_{a}^{\prime
}r_{b}^{\prime }\}$. We use approximation ($R>1,$ $I\ll 1$)%
\begin{equation}
\frac{1}{(e^{2\eta qd}-R)^{2}+I^{2}}\approx \frac{\pi }{I}\delta (e^{2\eta
qd}-R),
\end{equation}%
where $\delta (x)$ is the Dirac delta function. This leads to expression 
\begin{equation}
\mathcal{T}_{\mu \gamma }\approx e^{2\eta qd}\frac{\pi }{I}\frac{\delta
(\theta -\theta _{0})}{2\cos \theta _{0}yR(\theta _{0})}\sum\limits_{\mu }%
\mathcal{A}_{\mu \gamma },
\end{equation}%
where $\theta _{0}$ is defined by the equation $qd=\left( 2\eta _{0}\right)
^{-1}\ln R(\theta _{0})$ ($\eta _{0}=\sin \theta _{0}$). Now for $\mathcal{T}%
_{\gamma ,\theta <\theta _{th}}^{eff}$ we can write 
\begin{eqnarray}
\mathcal{T}_{\gamma ,\theta <\theta _{th}}^{eff} &=&\frac{15}{\pi ^{4}}%
\left\langle \int\limits_{0}^{q_{c}/q_{_{T}}}dx\frac{x^{3}}{e^{x}-1}\mathcal{%
T}_{\gamma }\right\rangle _{\theta <\theta _{th}}  \notag \\
&\approx &\dsum\limits_{\mu }\frac{15}{\pi ^{4}}\left\langle 
\begin{array}{c}
\int\limits_{0}^{q_{c}/q_{_{T}}}dx\frac{x^{3}}{e^{x}-1}\mathcal{A}_{\mu
\gamma }e^{2\eta qd} \\ 
\times \frac{\pi }{I}\frac{\delta (\theta -\theta _{0})}{2\cos \theta
_{0}yR(\theta _{0})}%
\end{array}%
\right\rangle _{\theta <\theta _{th}}  \notag \\
&=&\dsum\limits_{\mu }\frac{15}{\pi ^{4}}\frac{1}{dq_{T}}  \notag \\
&&\times \int\limits_{x_{th}}^{q_{c}/q_{_{T}}}dx\frac{x^{3}}{e^{x}-1}\frac{%
\pi \mathcal{A}_{\mu \gamma }(\theta _{0})}{I(\theta _{0})}\frac{\sin \theta
_{0}}{2\cos \theta _{0}}.
\end{eqnarray}%
Here $x_{th}=\ln R(\theta _{th})/(2q_{T}d\sin \theta _{th})$. Let us further
approximate (note that $\sin \theta _{0}=\ln (R)/2xdq_{T}$)%
\begin{equation}
\frac{\sin \theta _{0}}{\cos \theta _{0}}\approx \frac{\ln R(\theta _{0})}{%
2xdq_{T}},
\end{equation}%
when we obtain%
\begin{eqnarray}
\mathcal{T}_{\gamma ,\theta <\theta _{th}}^{eff} &\approx &\dsum\limits_{\mu
}\frac{15}{\pi ^{4}}\frac{1}{\left( 2dq_{T}\right) ^{2}}  \notag \\
&&\times \int\limits_{x_{th}}^{q_{c}/q_{_{T}}}dx\frac{x\pi f_{\mu \gamma
}(\theta _{0})\ln R(\theta _{0})}{e^{x}-1},
\end{eqnarray}%
where $f_{\mu \gamma }(\theta _{0})=\mathcal{A}_{\mu \gamma }(\theta
_{0})/I(\theta _{0}).$ Next we perform a Taylor series expansion for $%
f(\theta _{0})=\mathcal{A}_{\mu \gamma }(\theta _{0})/I(\theta _{0})$ close
to $\theta _{0}=0$:%
\begin{eqnarray}
f_{\mu \gamma }(\theta _{0}) &=&\mathcal{A}_{\mu \gamma }(\theta
_{0})/I(\theta _{0})\approx \theta _{0}^{n}\frac{f^{n}(0)}{n!}  \notag \\
&\approx &\left( \frac{\ln (R(0))}{2dq_{T}x}\right) ^{n}\frac{f_{\mu \gamma
}^{n}(0)}{n!}
\end{eqnarray}%
where $f^{n}(0)$ is the first derivative that is non-zero. We find%
\begin{eqnarray}
\mathcal{T}_{\gamma ,\theta <\theta _{th}}^{eff} &\approx &\dsum\limits_{\mu
}\frac{15}{\pi ^{3}}\frac{\left[ \ln R(0)\right] ^{n+1}}{\left(
2q_{T}d\right) ^{2+n}}  \notag \\
&&\times \frac{f_{\mu \gamma }^{n}(0)}{n!}F_{n}(\frac{q_{c}}{q_{_{T}}}%
,\theta _{th}),
\end{eqnarray}%
where $F_{n}(\frac{q_{c}}{q_{_{T}}},\theta
_{th})=\tint\nolimits_{x_{th}}^{q_{c}/q_{_{T}}}dx\frac{x^{1-n}}{\exp (x)-1}$.%
\newline

\subsection{Large angle contribution $\mathcal{T}_{\protect\gamma ,\protect%
\theta >\protect\theta _{th}}^{eff}$}

At large angles only long wave length phonons participate, i.e., the phonon
distribution can be approximated by $N(\omega _{\boldsymbol{q}},T)=[\exp
(\hbar \omega _{\boldsymbol{q}}/k_{B}T)-1]^{-1}\approx k_{B}T/\hbar \omega _{%
\boldsymbol{q}}$ and we find 
\begin{equation}
\mathcal{T}_{\gamma ,\theta >\theta _{th}}^{eff}\approx \frac{15}{\pi ^{4}}%
\frac{1}{\left( q_{T}d\right) ^{3}}\left\langle
\int\limits_{0}^{q_{c}d}dyy^{2}\mathcal{T}_{\gamma }(y,\theta )\right\rangle
_{\theta \geq \theta _{th}}.
\end{equation}


\input{dummy.tex}%
